\documentstyle[preprint,aps]{revtex}

\tighten

\begin{document}

\def\be{\begin{equation}}\def\ee#1{\label{#1}\end{equation}}
\def\bea#1\eea#2 {\begin{mathletters}\label{#2}\begin{eqnarray}#1\eea}
\def\eea{\end{eqnarray}\end{mathletters}}

\preprint{NSF-ITP-97-019  \qquad gr-qc/9703027}

\title{Black Hole Boundary Conditions and Coordinate Conditions}

\author{Douglas M. Eardley\footnote%
        {Electronic address: \tt doug@itp.ucsb.edu\hfil}}
\address{Institute for Theoretical Physics\\
        University of California\\
        Santa Barbara, CA 93106-4030}
\date{\today}

\maketitle

\begin{abstract}
This paper treats boundary conditions on black hole horizons for the full
3+1D Einstein equations.  Following a number of authors, the apparent
horizon is employed as the inner boundary on a space slice.  It is
emphasized that a further condition is necessary for the system to be
well posed; the ``prescribed curvature conditions" are therefore proposed
to complete the coordinate conditions at the black hole.  These conditions
lead to a system of two 2D elliptic differential equations on the inner
boundary  surface, which coexist nicely to the 3D equation for maximal
slicing (or related slicing conditions).  The overall 2D/3D system is
argued to be well posed and globally well behaved.  The importance of
``boundary conditions without boundary values" is emphasized.  This
paper is the first of a series.

\end{abstract}

\narrowtext


\section{Introduction}

If we need to define a flat surface, an elastic membrane --- such as
a soap bubble --- stretched
over a fixed rigid ring does a good job.  The shape of the
membrane is governed by the Laplace equation, with Dirichlet boundary
conditions at its edge determined by the ring.  What if we additionally
need the membrane to have a hole somewhere near its center?  If there
is a small rigid ring available, we can simply drop it onto the membrane,
and then cut out a hole in the membrane, fixing the new, inner edge to the 
small ring, again with Dirichlet boundary conditions.  Overall, the small
ring floats according to forces from the membrane, but the ring determines
the size and shape of the hole.  What do we do if no suitable small rigid
ring is available?  Our metaphor, admittedly loose, of course pertains to
the study of black holes by numerical relativity, where indeed there is
no suitable rigid ``ring".  

If we have closed loop of elastic string, it can serve as the inner
boundary. The size and shape of the hole is not fixed, but rather
determined by force balance between the string and the membrane.  We may
guess  the equilibrium shape of the hole to be a circle.  The curvature
of the circle, and hence the size of the hole, will be determined
by balance between tension along the string and surface tension
in the membrane.  The equilibrium will be described by some differential
equations:  In the membrane, the 2D Laplace equation; and along the
string, {\it two} stationary wave equations  --- two, because the string
has two transverse degrees of freedom in space.\footnote{The real string
also has a longitudinal degree of freedom; but this can be decoupled
by attaching the membrane to the string with a slip boundary ---
as with a soap bubble.}
Boundary conditions at the edge of the hole couple all of these
equations together.

Let us now pass from the metaphorical soap bubble to the real black hole.
Relativists have long hoped to pose numerical boundary conditions at
the horizon of a black hole, and use the horizon as the inner boundary
of the numerical grid.  Recently, progress has been achieved
in spherical symmetry by Scheel, Shapiro \& Teukolsky \cite{BD1,BD2}, and by
Anninos, Daues, Masso, Seidel, and Suen \cite{ADMSS} (see also
\cite{SS}).  These two groups carried
out important demonstrations of the feasibility of such horizon boundary
conditions by evolving spherically symmetric black holes (in Brans-Dicke
theory and Einstein theory respectively).  The basic idea is to make the
inner boundary an apparent horizon.

It would be highly desirable to extend such horizon boundary conditions
to more complex situations involving distorted or rotating black
holes, and black hole binaries.
The main purpose of this paper is to work out the underlying theory of
these coordinate conditions in 3D generality, as a step of such
an extension.  In particular the 2D differential equations governing such
inner boundary surfaces will be studied with a view toward making them
well-posed, when connected to the 3D equation for maximal slicing in
relativity, or to various other slicing conditions ({\it i.e.,} choice
of lapse) such as generalizations of maximal slicing \cite{BDSSTW}\@.
The basic idea of using an apparent horizon will turn out to work, but
several difficult issues crop up along the way, and must be dealt with.
We need boundary conditions at the horizon, but we do not know how to
supply boundary values there;  how to resolve this issue of ``boundary
conditions without boundary values" is explained along the way.

The development in this paper will not make specific assumptions
about the conditions for 3+1D slicing and 3D spatial coordinates,
though we will often assume that these conditions are implemented by
some kind of 3D elliptic or parabolic equations, which will require
boundary conditions at the hole.  Hyperbolic operators ({\it e.g.},
arising in harmonic slicing or harmonic coordinates) form quite a
different case, and probably a simpler one, since they do not require
boundary conditions at the hole.\footnote{On the other hand, it is
not clear whether harmonic slicing avoids coordinate singularities.}
(I am grateful to Greg Cook for pointing this out.)

In future papers in this series, further topics will be treated:
\begin{itemize}
  \item Horizon boundary conditions for stationary black holes,
	relevant to the late stages of numerical calculations.
  \item	The connection between spatial coordinate conditions,
	{\it i.e.,} choice of shift, and horizon boundary conditions;
	this is especially important for rotating black holes.
  \item Boundary conditions for the constraint equations of general
	relativity on horizons; again a key issue is
	``boundary conditions without boundary values".
\end{itemize}

\section{Most Apparent Horizons are Wild}

Hawking \cite{HE} defined trapped surfaces and apparent horizon as follows.

\proclaim Definition 1.  A trapped surface $S$ is an achronal 2-surface
in spacetime for which the outward null convergence obeys
\be
	\rho > 0
\ee{traps}
A marginally trapped surface $S$ is one for which
the outward null convergence obeys
\be
	\rho = 0
\ee{mtraps}

The basic property of these surface is:

\proclaim Proposition 2. (Hawking \& Ellis \cite{HE}).  Given a trapped or
marginally trapped surface $S$.  No point $p$ on $S$ can lie outside the
event horizon.

\proclaim Definition 3. In a given space slice $V$, an apparent horizon
is the outward boundary of trapped surfaces that lie in $V$\@.

Notice that the given slice $V$ is
an essential ingredient of the definition.  To say that ``We are
going to choose a slice that meets an apparent horizon..."
is a {\it circular definition}.  This issue is not just a niceity;
to ignore it can lead to serious trouble, as we will soon see.
One way to see the problem is to observe that {\it any} slice
passing far enough into the black hole --- over a wide range ---
will meet an apparent horizon, and therefore will obey the
boundary condition.  A condition that excludes nothing is
not a useful condition.

To de-circularize the definitions, we must drop the slice, and
refer not to an apparent horizon, but to a marginally trapped
surface.
So now we can sensibly say ``We are going to choose a slice that
meets a marginally trapped surface."  And we can proceed to build
a 3+1D code that uses marginally trapped surfaces $S$ as the inner
boundary of the slices.  However, {\it this code may not
work; it is vulnerable to crashing after a short time.}
The pitfall is that ``most" marginally trapped surfaces are wild
surfaces;  this is very unlike the familiar situation with maximal
slices, which are automatically smooth (technically, thanks to
elliptic regularity).  The wildness of marginally trapped surfaces
is intrinsically a non-spherically-symmetric phenomenon, and does
not show up at all in spherically symmetric setting --- consistent
with the success of Anninos, Daues, Masso, Seidel, and Suen.

A related property of a generic marginally trapped surface, is that it
can always be deformed spatially outward from any point on it:

\proclaim Proposition 4.  Given any smooth marginally trapped surface
$S$, such that either $\sigma$ or $T_{\mu\nu}l^\mu l^\nu$ does not vanish
identically on $S$\@.  Given any point $p$ on $S$, and given a spacelike
outward-pointing
vector $u$ at $p$\@.  Then $S$ can be locally perturbed into a 1-parameter
family of marginally trapped surfaces $S(\epsilon)$ so that $p$ moves
in the $u$-direction.

(Here $\sigma$ is the outward null shear, $T_{\mu\nu}l^\mu l^\nu$ is the
outward null component of the stress energy tensor, and we will assume
throughout that $T_{\mu\nu}l^\mu l^\nu$ obeys the dominant energy condition.
Actually Proposition 4 is true for any $u$, but is most interesting
if $u$ is spacelike outward-pointing.)

This leads to a puzzle.  Start with an apparent horizon $S$ and some
point $p$ on it.  What is to keep us from continually perturbing it,
extending the 1-parameter family $S(\epsilon)$, until $p(\epsilon)$
passes outside the event horizon, contradicting Proposition 2?  The
only thing that can go wrong is that $S$ must ``go wild" ---
{\it i.e.,} cease being smooth --- first.

In fact, the following conjecture, at first a bit startling, can be
surmised as the obvious general answer to the puzzle.

\proclaim Conjecture A. The outward boundary in spacetime of marginally
trapped surfaces is the event horizon.

I have proven this conjecture under some assumptions, plausible
but not rigorously established, about how spacetime settles down
to a nonextremal black hole subsequent to gravitational collapse.
The trapped surfaces $S$ that pass close to the event horizon are
wild.  Let us give an example to illustrate what happens.

Sometime during the dynamical phase of collapse, after the black
hole has formed but before it has settled down, choose a slice
that contains an apparent horizon $S$, and choose a point $p$ on
$S$.  Also choose any point $q$ on the intersection of the slice
with event horizon.  In the usual picture, $q$ is considerably
outside the apparent horizon, so it is unclear how any trapped
surface can pass close to $q$\@.)  However, by Proposition 4,
we can now extend a 1-parameter family of marginally trapped
surfaces $S(\epsilon)$, so that $p(\epsilon)$ remains in the
slice $V$ (though most of $S(\epsilon)$ does not) and so that
$p(\epsilon)\rightarrow q$ at some parameter value $\epsilon_0$\@.
What happens is that all of $S(\epsilon)$ approaches the event
horizon as $\epsilon\rightarrow\epsilon_0$, but that most of
$S(\epsilon)$ moves far to the future.
Therefore the trapped surfaces $S$ which pass close to $q$ look as
follows.  Most of $S$ lies far to the future, very close to the
event horizon of the settled-down black hole.  Only a thin tendril
of $S$ extends back near $q$, in a thin tubular neighborhood of a
generator of the event horizon.  Though this tendril is
nearly null, it is still part of a spacelike surface.

Let us see what this means for numerical relativity.  Each marginally
trapped surface $S(\epsilon)$ can serve as the inner boundary of
a slice $V(\epsilon)$, say a maximal slice, extending to spatial
infinity.  Moreover we can arrange for $\epsilon$ to be proper time
at infinity, so that the slicing ``goes wild" at finite time.
This is the serious trouble that such a code can run into.

Details will be published elsewhere.

\section{The Prescribed Convergence Conditions for 2-Surfaces}

This is not to say that apparent horizons are necessarily a bad idea
for inner boundary conditions, only that an ingredient is missing.
We propose that the missing ingredient is the simplest possible thing,
the convergence $\rho'$ of the inward null normal to a surface $S$,
and that the well posed way to put an inner boundary on a slice is
to prescribe both the outward null convergence $\rho$ and the inward
null convergence $\rho'$.

\subsection{The Two Mean Extrinsic Curvatures of a 2-Surface in Spacetime}

Consider a spatial 2-surface $S$ immersed in spacetime.  (Throughout, $S$
will be topologically a 2-sphere unless otherwise noted.)  All over $S$,
we can choose an orthonormal frame of reference ${{\bf e}_{\hat\alpha}}$
so that the time axis ${{\bf e}_{\hat0}}$ and one of the spatial axes
${{\bf e}_{\hat1}}$ are normal to $S$, while the other two spatial axes
${{\bf e}_{\hat2}}$, ${{\bf e}_{\hat3}}$ are tangent to $S$.  Then $S$
has defined on it two {\it mean extrinsic curvatures,} namely $H_0$ in
the $\hat0$-direction, and $H_{\hat1}$ in the $\hat1$-direction.  If
$\sqrt{h}$ is the element of area on $S$, then in suitable local
coordinates we can take as the definitions
\bea
	\partial_{\hat0} \sqrt{h} &=& -H_{\hat0}\sqrt{h}	\\
	\partial_{\hat1} \sqrt{h} &=& -H_{\hat1}\sqrt{h}
\eea{defH01}
(the minus sign in these equations is a matter of convention).  In any
coordinate system, these mean extrinsic curvatures form a 4-vector
$H_\mu$ orthogonal to $S$.

Alternatively we can follow null methodology \cite{PR} and choose
two orthonormal null vectors, the outward and inward null normals to
$S$ respectively, as
\bea
	{\bf l} &=& \frac{{{\bf e}_{\hat0}}+{{\bf e}_{\hat1}}}{\sqrt2}	\\
	{\bf n} &=& \frac{{{\bf e}_{\hat0}}-{{\bf e}_{\hat1}}}{\sqrt2}
\eea{defln}
with $l_\mu l^\mu = 0 = n_\mu n^\mu$, $l_\mu n^\mu = -1$, and then
the null convergences of ${\bf l}$ and ${\bf n}$ are
\bea
	\rho &=& \frac{H_{\hat0} + H_{\hat1}}{\sqrt2}		\\
	\rho' &=& \frac{H_{\hat0}-H_{\hat1}}{\sqrt2}
\eea{defrho}

Notice, however, that we had to choose some frame of reference on $S$
to define $H_{\hat0}$ and $H_{\hat1}$ (or $\rho$ and $\rho'$).  The
arbitrariness in choice of frame amounts to a boost in the 1-direction
(a {\it GHP boost}) all over $S$ \cite{PR}\@.  Under such a boost these
quantities transform like
\bea
     {\bf l} \rightarrow \tilde{\bf l}	&=& e^\psi {\bf l}	\\
     {\bf n} \rightarrow \tilde{\bf n}	&=& e^{-\psi} {\bf n}	\\
	\rho \rightarrow \tilde\rho	&=& e^\psi \rho		\\
       \rho' \rightarrow \tilde\rho'	&=& e^{-\psi} \rho'
\eea{boostln}
where the boost rapidity $\psi$ is an arbitrary function on $S$.
Only one boost-invariant scalar can be formed from the mean extrinsic
curvatures, namely $H_\mu H^\mu = -H_{\hat0}^2+H_{\hat1}^2  =
-2\rho\rho'$.

How then do we determine a 2-surface $S$ in spacetime?  Since $S$
has two transverse degrees of freedom, we need two conditions at
each point of $S$, that is, two equations in two coordinates.
It is tempting therefore to put conditions on the two quantities
$\{H_{\hat0},H_{\hat1}\}$ or equivalently $\{\rho,\rho'\}$:
\bea
	\rho = \hbox{(some fixed function on $S$)}		\\
	\rho' = \hbox{(some other fixed function on $S$)}
\eea{fixrho}
(For instance, to use an apparent horizon we can just take $\rho=0$
as one of the two conditions.)  These equations will be called the
{\it equations of prescribed convergence}, or, for short, {\it the
PC equations.}  However the PC equations are not enough by themselves,
because of the boost arbitrariness, Eq.~(\ref{boostln}).  We are
missing one condition, namely something to fix $\psi$ as a function
of two coordinates on $S$.

If we have already decided on a slicing condition for spacetime,
{\it e.g.,} maximal slicing, or something related to it, then the
slice itself provides the missing condition:  We can use the unit
timelike normal to the slice to fix the $\hat0$-direction at $S$.
Then the PC equations, together with the maximal slicing equation,
are just enough.  This is of course no guarantee that the equations
form a {\it well-posed} system, meaning that $V$ and $S$ can never
``go wild".   However it is a reasonable conjecture that this is
so, and this paper will present a considerable amount of evidence
in favor of this conjecture.

\subsection{The 3+1 and 2+1 Splits}

The spacetime metric is
\be
	ds^2 = -(\alpha^2 - \beta_i\beta^i)dt^2 + 2\beta_i dx^i dt +
		\gamma_{ij}dx^idx^j
\ee{4metric}

The notation is now necessarily going to become a little complicated,
so the reader is asked to be patient.   Starting from the full spacetime
geometry, we choose slices and carry out the standard 3+1 split.  The
unit future-pointing timelike normal to the slices is denoted
${{\bf e}_{\hat0}}$.  Spacetime indices run over $\mu,\nu,\ldots=0,1,2,3$,
where $0$ denotes time $t$;  spatial indices run over $i,j,\ldots=1,2,3$,
and can be lowered and raised with the spatial metric $\gamma_{ij}$
and its inverse $\gamma^{ij}$.

Each space slice $t={\rm const}$ has an inner boundary on a 2-surface $S$.
For convenience in this paper,
we will throughout choose spatial coordinates $\{x^i\}$ and shift vector
$\beta^i$ so that
\be
\hbox{$S$ always lies at } r=r_0={\rm const}, \hbox{ where } r\equiv x^1;
\ee{rcon}
This means that $\beta_i$ is not freely specifiable at $S$,
but must be chosen to match the motion of $S$:
\be
	B \equiv \beta_i{e^i_{\hat1}} \quad\hbox{fixed at $S$ by the
							PC conditions}
\ee{betacon}
here $e^i_{\hat1}$ is the unit outward spatial normal to $S$.
One would actually like more general spatial coordinates, to allow the
black hole to move through the spatial coordinate system --- to ``fly through
the grid".  The generalization to such coordinates is straightforward
but is not developed in this paper.

The Riemannian 2-metric on $S$ will be denoted $h_{ab}$;  surface
indices run over $a,b,\ldots=2,3$ and can be lowered and raised
with $h_{ab}$ and its inverse $h^{ab}$.  We use a 2+1 split at
$S$, and 2-tensors on $S$ will carry indices $a,b,\ldots$.
The spatial extrinsic 2-curvature of $S$
with respect to the spatial normal direction ${e^i_{\hat1}}$ is a 2-tensor
\be
	H_{ab} \equiv -\frac12 {\cal L}_{\hat1} h_{ab}
\ee{defHab}
where $\cal L$ denotes Lie derivative,
and the spatial mean extrinsic 2-curvature $H_{\hat1}$ is its 2-trace,
\be
  H_{\hat1}	= h^{ab}H_{ab}
		= - \frac1{\sqrt h}\partial_{\hat1}\sqrt h
\ee{defH}
where $h\equiv {\rm det}h_{ab}$.
The Gauss-Codazzi equations of the 2+1 split imply
\bea
   {}^3\!R	&=& 2\partial_{\hat1}H_{\hat1} - H_{\hat1}^{\,2}
		- H_{ab}H^{ab} + \,{}^2\!R		\\
   h^{ab}\,{}^3\!R_{ab}	&=& \partial_{\hat1}H_{\hat1}
		- H_{ab}H^{ab} + \,{}^2\!R		\\
   {}^2\!D_aK^a_{\hat1} &=& \partial_{\hat1}H_{\hat0}
		- H_{\hat1}H_{\hat0} + H_{\hat1}{\rm tr}K
\eea{GC2}
where ${}^2R_{ab}$ is the Ricci 2-tensor of $h_{ab}$ and
${}^2R=h^{ab}R_{ab}$.

We also need the timelike extrinsic 2-curvature of $S$ with respect to the
timelike normal direction ${{\bf e}_{\hat0}}$; it is the 2-tensor gotten
by 2+1 projection of $K_{ij}$,
\be
	J_{ab} = \perp K_{ab}
\ee{defJab}
and the timelike mean extrinsic 2-curvature $H_{\hat0}$ is its 2-trace,
\be
        H_{\hat0} = h^{ab}J_{ab} = h^{ab}K_{ab};
\ee{defJ}
here $\perp$ denotes the projection of 3-tensors into 2-tensors
at $S$.

\subsection{Evolution of the Mean Extrinsic Curvatures}

We will present a system of equations which is fully covariant
under transformations of the spatial coordinates ${x^i}$ and shift
$\beta^i$ that preserve the inner-boundary constraints at $S$,
Eqs.~(\ref{rcon}, \ref{betacon}).  That is, we shall derive a system of
equations which can be used with any choice of spatial gauge away from
$S$.  Some intermediate calculations, though, are most easily done
in a particular spatial coordinate system, namely, Gaussian normal
coordinates in space.
Thus our spatial metric can be taken as
\be
	\gamma_{ij}dx^idx^j = dr^2 +h_{ab}dx^adx^b
\ee{3metric}
where $r\equiv x^1$.  The inner boundary in spacetime is the hypersurface
$r=r_S={\rm const}$, consisting of all the $S$ for all $t$, which we want
to be a spacelike or null hypersurface lying within or on the event horizon.
Values at $S$ will be denoted "$\big|_S$".  Thus by Eq.~(\ref{betacon})
\bea
	\beta_{\hat1}\big|_S &\equiv& B			\\
\noalign{\hbox{and also we will denote the inner boundary value the lapse
at $S$ as}}
	\alpha\big|_S &\equiv& A			\\
\noalign{\hbox{and its spatial normal derivative as}}
	\partial_{\hat1}\alpha\big|_S \equiv A_{\hat1}
\eea{bdybaa}

At the inner boundary surface $S$, the two mean curvatures
$H_{\hat0}\big|_S$ and $H_{\hat1}\big|_S$ are the quantities we wish
to prescribe.  Then we wish to derive these inner boundary values
for the lapse $\alpha$ and shift $\beta^i$.

From these definitions and the Einstein equations follow the evolution
equations for $H_{\hat0}$ and $H_{\hat1}$:
\def\pnt{\partial_t^{(n)}}
\bea
   \pnt H_{\hat0} &=&	\biggl[\frac12(H_{\hat0}^2-H_{\hat1}^2)
		+ \frac12(H_{ab}H^{ab}+J_{ab}J^{ab}) -  K_r^aK_a^r +
		8\pi T_{\hat0\hat0}
		+\frac12\,{}^2\!R - \,{}^2\!\Delta\biggr]A	\nonumber\\
     &&\mbox{}	+ \biggl[J_{ab}H^{ab} + H_{\hat0}H_{\hat1}
		- H_{\hat1}{\rm tr}K - 8\pi T_{\hat0r}
		+ ({}^2\!D_aK_r^a) + 2K_r^a\,{}^2\!D_a\biggr]B	\nonumber\\
     &&\mbox{}	+ \biggl[H_{\hat1}\biggr]A_{\hat1}		\\
   \pnt H_{\hat1} &=& \biggl[J_{ab}H^{ab} - 8\pi T_{\hat0r}
		- ({}^2\!D_aK_r^a) - 2K_r^a\,{}^2\!D_a\biggr]A	\nonumber\\
     &&\mbox{}	+ \biggl[\frac12(H_{\hat0}^2+H_{\hat1}^2)
		+ \frac12(H_{ab}H^{ab}+J_{ab}J^{ab})
		+ K_r^aK_a^r - H_{\hat0}{\rm tr}K		\nonumber\\
     &&\mbox{}	+ 8\pi T_{\hat0\hat0}
		- \frac12\,{}^2\!R + \,{}^2\!\Delta\biggr]B	\nonumber\\
     &&\mbox{}	+ \biggl[H_{\hat0}\biggr]A_{\hat1}
\eea{evolH}
where $\pnt$ denotes the projection of the time evolution operator
$\partial_t\equiv\partial/\partial t$ normal to $S$,
\be
  \pnt \equiv	A e_{\hat0}^\mu{\partial\over\partial x^\mu} +
		B e_{\hat1}^\mu{\partial\over\partial x^\mu}
\ee{defpnt}
These equations are valid in any 3+1 coordinate system that obeys
Eqs.~(\ref{rcon}, \ref{betacon}) at $S$.

If we now view $H_{\hat0}$ and $H_{\hat1}$ as prescribed functions of
$(t,x^a)$, Eqs.~(\ref{evolH}) become a set of two 2nd-order coupled
partial differential equations in $\{x^a\}$ on $S$ for three unknown
functions $A$, $B$ and $A_{\hat1}$.  Thus, the inner boundary conditions
for the lapse $\alpha$ and the radial component of the shift
$\beta^{\hat1}$are fixed.  The tangential boundary value of the shift
is still free, and can be chosen separately to enforce, say, ``slip" or
``no-slip" conditions as desired.

\subsection{The Naturality of the Inner Boundary Conditions}

It may seem surprising that these equations involve $A_{\hat1}$ as well
as $A$ and $B$;  however a little reflection shows this to be natural
and in fact desirable.  To illustrate this, restrict to maximal slicing
${\rm tr}K=0$ temporarily.   The maximal slicing equation
\be
	0 = ({}^3\Delta - K_{ij}K^{ij} - matter)\alpha
\ee{maxslice}
is a 2nd-order elliptic equation in $(x^i)$ for $\alpha$, and therefore
admits at the inner boundary $S$ any of these well-posed boundary
conditions:
\bea
  \alpha\big|_S &=& A
	\hbox{\quad Dirichlet boundary conditions, or}\\
  \partial_{\hat1}\alpha\big|_S &=& A_{\hat1} \hskip-0.40em
	\hbox{\quad Neumann boundary conditions, or more generally,}\\
  F_D\alpha\big|_S + F_N\partial_{\hat1}\alpha\big|_S &=& 0
	\hbox{\quad homogeneous mixed boundary conditions;}\\
  F_D\alpha\big|_S + F_N\partial_{\hat1}\alpha\big|_S &=& F
	\hbox{\quad inhomogeneous mixed boundary conditions;}
\eea{specond}
where $A$, $A_{\hat1}$, $F_D$, $F_N$, and $F$ are prescribed functions
of $(t,x^a)$.

Conditions (\ref{specond}c,\ref{specond}d) simply say that there
exists a prescribed linear relation between $\alpha$ and $\alpha_{\hat1}$
on $S$, but that neither is fixed individually;  the distinction between
these conditions (\ref{specond}c,\ref{specond}d) is the function $F$ on
the right-hand-side, which may be either 0 (homogeneous case) or a given
function (inhomogeneous case).

Now, given $H_{\hat0}$ and $H_{\hat1}$,
Eqs.~(\ref{evolH}) do give two prescribed linear relations among the three
functions $(B-A)$, $(B+A)$ and $A_{\hat1}$, albeit implicit ones.  One
can imagine using one of Eqs.~(\ref{evolH}) to eliminate $B$, whereupon
the other becomes a single linear relation, in the form of
Eq.~(\ref{specond}c), between $A$ and $A_{\hat1}$ --- albeit an implicit
one, involving Green functions of the operators ${\cal D}$ and ${\cal D}'$.
Thus, Eqs.~(\ref{evolH})
{\it do} appear to give well-posed boundary conditions for maximal slicing,
Eq.~(\ref{maxslice}), and they also should do so for attractive
generalizations of maximal slicing \cite{ADMSS}.

In fact, not only is it admissible to have homogeneous mixed boundary
conditions for the lapse $\alpha$ at $S$, it is also desirable.  Imagine
we instead used Dirichlet boundary conditions (\ref{bdybaa}a).  How
big do we then make the inner  boundary value $A$? --- by which we mean,
what fixes the overall scale of $A$ at the inner boundary?
We would like to make it big enough so that, over a long
evolution, the inner boundary just ``keeps up" with the outer boundary,
neither shooting way ahead, or falling way behind.  How big is that?
The first guess is $A=1$, but that cannot be right, because the inner
boundary is in the strong field region.   The lapse equation is not
going to tell us how big to make $A$, precisely because it is happy
with any  $A$.  Thus a pure Dirichlet boundary condition is not
generally going to work well for long evolutions.

Thus, mixed homogeneous boundary conditions resolve this issue.
They provide ``boundary conditions without boundary values".

A homogeneous mixed boundary condition can work well --- it can be
thought of as a ``feedback mechanism".  If the inner boundary falls
behind the outer boundary, then $A_{\hat1}$ will become large and
positive, and the mixed boundary condition then can ``tell" $A$
to become larger.  If the inner boundary shoots ahead of the outer
boundary, then $A_{\hat1}$ will become large and negative, and the
mixed boundary condition then can ``tell" $A$ to become smaller.  A
concrete example of this ``mixed-boundary-condition feedback mechanism"
will be presented below.   Since the boundary condition is homogeneous,
no data need be given at the inner boundary to fix the overall
scale of $A$;  this scale is self-adjusting.

\subsection{The PC Equations -- Inner Boundary Conditions for Prescribed
$\rho$ and $\rho'$}

From Eqs.~(\ref{defrho},\ref{evolH}) can be found the evolution equations
for the null convergences $\rho$ and $\rho'$.  When we take $\rho$ and
$\rho'$ as prescribed functions of $(t,x^a)$, these equations become
the {\it equations of prescribed convergence}, or, for short, {\it the
PC equations:}
\bea
 \pnt\rho	&=& \frac1{\sqrt2}\biggl[ \rho( \rho' + 2\rho
		- {\rm tr}K/\sqrt2) + \frac12{\cal E}
					\biggr](B+A)	\nonumber\\
     &&\mbox{}	+   \frac1{\sqrt2}\biggl[ \rho(-\rho' +  \rho
		- {\rm tr}K/\sqrt2) + \frac12{\cal D}
		 			\biggr](B-A)	\nonumber\\
     &&\mbox{}	+ \biggl[\rho\biggr]A_{\hat1}			\\
 \pnt\rho' &=& -\frac1{\sqrt2}\biggl[ \rho'(-\rho +  \rho'
		- {\rm tr}K/\sqrt2) + \frac12{\cal D}'
					\biggr](B+A)	\nonumber\\
     &&\mbox{}	-     \frac1{\sqrt2}\biggl[ \rho'( \rho + 2\rho'
		- {\rm tr}K/\sqrt2) + \frac12{\cal E}'
		 			\biggr](B-A)	\nonumber\\
     &&\mbox{}	- \biggl[\rho'\biggr]A_{\hat1}
\eea{evolrho}
a set of two 2D elliptic equations relating three unknown functions
$A$, $B$, $A_{\hat1}$, where
\bea
  {\cal D} &\equiv&
	h^{ab}({}^2\!D_a+K_a^r)({}^2\!D_b+K_b^r) - \frac12\,{}^2\!R	\\
  {\cal D}' &\equiv&
	h^{ab}({}^2\!D_a-K_a^r)({}^2\!D_b-K_b^r) - \frac12\,{}^2\!R	\\
  {\cal E} &\equiv&
	\frac12(J_{ab}-H_{ab})(J^{ab}-H^{ab}) - 2\rho^2
	+ 8\pi(T_{\hat0\hat0}-T_{\hat0r})				\\
  {\cal E}' &\equiv&
	\frac12(J_{ab}+H_{ab})(J^{ab}+H^{ab}) - 2{\rho'}^2
	+ 8\pi(T_{\hat0\hat0}+T_{\hat0r})
\eea{defDE}
The latter quantities obey some useful relations.  The two differential
operators ${\cal D}$ and ${\cal D}'$ live on $S$; they are not
generally self-adjoint, due to the terms in $K_a^r$, but are the adjoints
of each other:
\be
	D^\dagger = D'.
\ee{adjD}
The two scalar functions ${\cal E}$ and ${\cal E}'$ are nonnegative:
\bea
   {\cal E}	&\ge& 0		\\
   {\cal E}'	&\ge& 0
\eea{posE}
as long as matter obeys the dominant energy condition, which we will
assume throughout.
These equations are valid in any 3+1 coordinate system that obeys
Eqs.~(\ref{rcon}, \ref{betacon}) at $S$.

\subsection{Special Case: Maximal Slices Bounded by Apparent Horizons}

As a special but important choice of coordinate conditions --- still
general enough to study binary black hole coalescences in 3+1D ---
let us take:
\bea
 {\rm tr}K	&=& 0	\hbox{\qquad(maximal slicing)}\\
    \rho	&=& 0	\hbox{\qquad(apparent-horizon inner boundary)}\\
   \rho'	&=& F(x^a) \hbox{\qquad(some prescribed function)}
\eea{specase}

Then the PC equations reduce to
\bea
  0	&=&    {\cal E}(B+A) + {\cal D}(B-A)		\\
  0	&=&   ({\cal D}' + 2{\rho'}^2) (B+A)
	    + ({\cal E}' + 4{\rho'}^2) (B-A) + 2\sqrt2\rho' A_{\hat1}
\eea{specrho}

\subsection{Special Case: Stationary Black Holes}

Stationary black holes are relevant because calculations of binary
coalescence will eventually settle down to a stationary black hole,
and coordinate conditions are desirable that will somehow ``lock onto"
the geometry of the stationary black hole and render it recognizable.
For a stationary black hole, we have
\bea
	\rho		&=& 0			\\
	{\cal E}	&=& 0
\eea{statbh}
and then from Eq.~(\ref{defDE}a) we have immediately on $S$
\be
	B = A
\ee{BeqA}
so that the PC conditions reduce to one 2D elliptic equation
relating two unknown functions $A$, $A_{\hat1}$:
\bea
 \pnt\rho' &=& -\sqrt2\biggl[ \rho'(\rho'
		- {\rm tr}K/\sqrt2) + \frac12{\cal D}'
		\biggr] A - \biggl[\rho'\biggr]A_{\hat1}.
\eea{statDE}
The operator on $A$ is not obviously invertible;  however in a later
paper in this series, it will be proven that this operator is in fact
invertable under fairly general conditions for a nonmaximal stationary
black hole, and so the PC conditions can be expected to successfully
``lock onto" the Kerr black hole at late times.

\section{Conclusion}

We have studied the apparent horizon boundary condition $\rho=0$
on a spacelike 2-surface $S$ as an inner coordinate condition at a
black hole.  Since most apparent horizons are wild, another condition
is required to ensure a well posed evolution.  We propose the {\it
Prescribed Curvature Equations,} or for short the {\it PC Equations}
\bea
	\rho = \hbox{(some fixed function on $S$)}		\\
	\rho' = \hbox{(some other fixed function on $S$)}
\eea
as an effective condition at the black hole.  These equations
have the following properties:
\begin{itemize}
  \item	They give rise to a system of two linear 2D elliptic equations
	on $S$ for three unknown boundary values of the lapse and shift.
  \item They therefore connect well with maximal slicing, and with
	related slicing conditions that involve 3D elliptic or
	parabolic equations.
  \item They are well posed and solvable for stationary black holes.
  \item They appear likely to be well posed and solvable under very
	general conditions.
\end{itemize}

Some additional numerical work will be required to solve the PC
equations as part of a numerical relativity code.  However, solving
these equations cannot be not much harder than finding apparent horizons
in the first place, (see, {\it e.g.,} \cite{AH3,ACLMSS}) and may well be
easier because these equations are linear.  Therefore the extra work
seems unlikely to be prohibitive.

\acknowledgements

This research was supported in part by the National Science Foundation
under Grant Nos.~PHY94-07194 and PHY90-08502 at ITP and UCSB\@.  I am
grateful for the hospitality of the the Texas/Los Alamos Workshop (IGPP,
Feb.~1997), and grateful to workshop participants for many helpful
comments on a version of this work.  I am also grateful to Greg Cook,
Sean Hayward, and Wai-Mo Suen for comments on the manuscript.


\begin{references}
\frenchspacing

\bibitem{BD1} M.A. Scheel, S.L. Shapiro, and S.A. Teukolsky,
{\sl Collapse to black holes in Brans-Dicke theory.  I. Horizon boundary
conditions for dynamical spacetimes,} gr-qc/9411025,
{\it Phys.\ Rev.} {\bf D51}, 4208 (1995).

\bibitem{BD2} M.A. Scheel, S.L. Shapiro, and S.A. Teukolsky,
{\sl Collapse to black holes in Brans-Dicke theory.  II. Comparison with
general relativity,} gr-qc/9411026,
{\it Phys.\ Rev.} {\bf D51}, 4236 (1995).

\bibitem{ADMSS} P. Anninos, G. Daues, J. Mass\'o, E. Seidel, and W. Suen,
{\sl Horizon boundary condition for black hole spacetimes,}
gr-qc/9412069,
{\it Phys.\ Rev.} {\bf D51}, 5562 (1995).

\bibitem{SS} E. Seidel and W. Suen,
{\sl Towards a singularity-proof scheme in numerical relativity,}
{\it Phys.\ Rev.\ Lett.} {\bf 69}, 1845 (1992).

\bibitem{BDSSTW} J. Balakrishna, G. Daues, E. Seidel, W. Suen, M. Tobias,
and E. Wang
{\sl Coordinate Conditions and Their Implementation in 3D Numerical
Relativity,} gr-qc/9601027,
{\it Class.\ Quant.\ Grav.} {\bf13}, L135-L142 (1996).

\bibitem{HE} S.W. Hawking \& G.F.R. Ellis, {\it The Large Scale Structure of
Space-Time} (Cambridge University Press, Cambridge, 1973) p.~320.

\bibitem{PR} R. Penrose and W. Rindler, {\it Spinors and Space-Time. Vol.~I}
(Cambridge University Press, Cambridge, 1984) Sect.~4.14.


\bibitem{AH3} T.W. Baumgarte, G.B. Cook, M.A. Scheel, S.L. Shapiro,
and S.A. Teukolsky,
{\sl Implementing an apparent-horizon finder in three dimensions,}
gr-qc/9606010,
{\it Phys.\ Rev.} {\bf D54}, 4849 (1996).

\bibitem{ACLMSS}
P. Anninos, K. Camarda, J. Libson, J. Mass\'o, E. Seidel, and W. Suen
{\sl Finding Apparent Horizons in Dynamic 3D Numerical Spacetimes,}
gr-qc/9609059,
submitted to {\it Phys.\ Rev.\ D}.

\end{references}
\end{document}